\documentclass[12pt]{iopart}

\usepackage{iopams}  
\usepackage{epsfig}
\begin{document}

\title{Fluctuation Effects on $R_{pA}$ at High Energy}

\author{Misha~Kozlov$^a$, ~Arif~I.~Shoshi$^a$ ~and ~Bo-Wen Xiao$^b$}
\address{$^a$ ~Fakult{\"a}t f{\"u}r Physik, Universit{\"a}t Bielefeld, D-33501 Bielefeld,Germany}
\address{$^b$ ~Department of Physics, Columbia University, New York, NY, 10027, USA}
%
%
\begin{abstract}
  We discuss a new physical phenomenon for $R_{pA}$ in the fixed coupling case, 
  the total gluon shadowing, which arises due to the effect of gluon number
  fluctuations.
\end{abstract}

We study the ratio of the unintegrated gluon distribution of a nucleus
$h_A(k_{\perp},Y)$ over the unintegrated gluon distribution of a proton
$h_p(k_{\perp},Y)$ scaled up by $A^{1/3}$
\begin{equation}
R_{pA} = \frac{h_{A}\left( k_{\perp },Y\right) }{A^{\frac{1}{3}}\ h_{p}\left(
    k_{\perp },Y\right) } \ .
\label{R_pA}
\end{equation}
This ratio is a measure of the number of particles produced in a
proton-nucleus collision versus the number of particles in proton-proton
collisions times the number of collisions. The transverse
momentum of gluons is denoted by $k_{\perp}$ and the rapidity variable by $Y$.

\begin{figure}[htb]
\setlength{\unitlength}{1.cm}
\par
\begin{center}
\epsfig{file=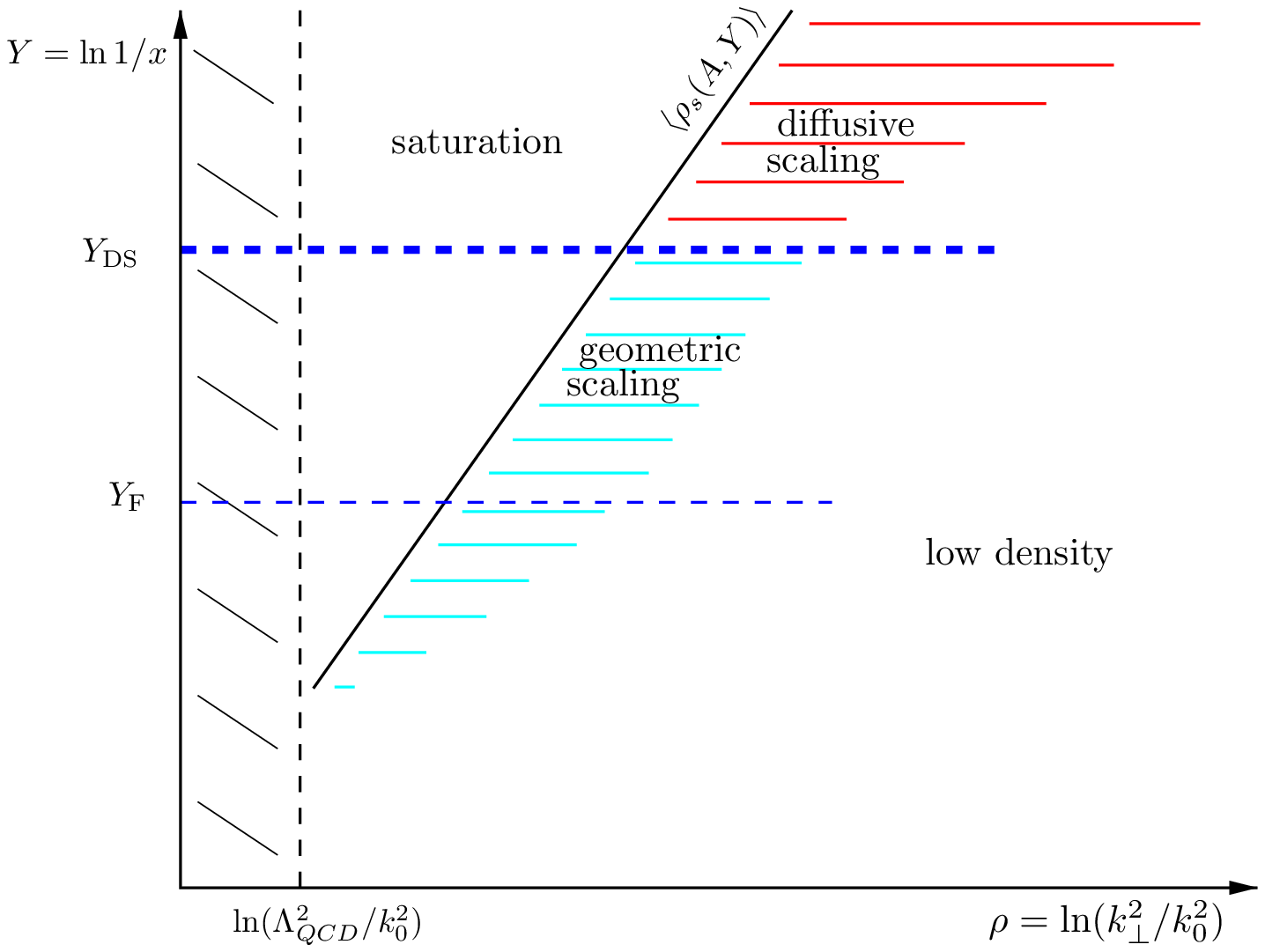, width=7.5cm} \hfill
\epsfig{file=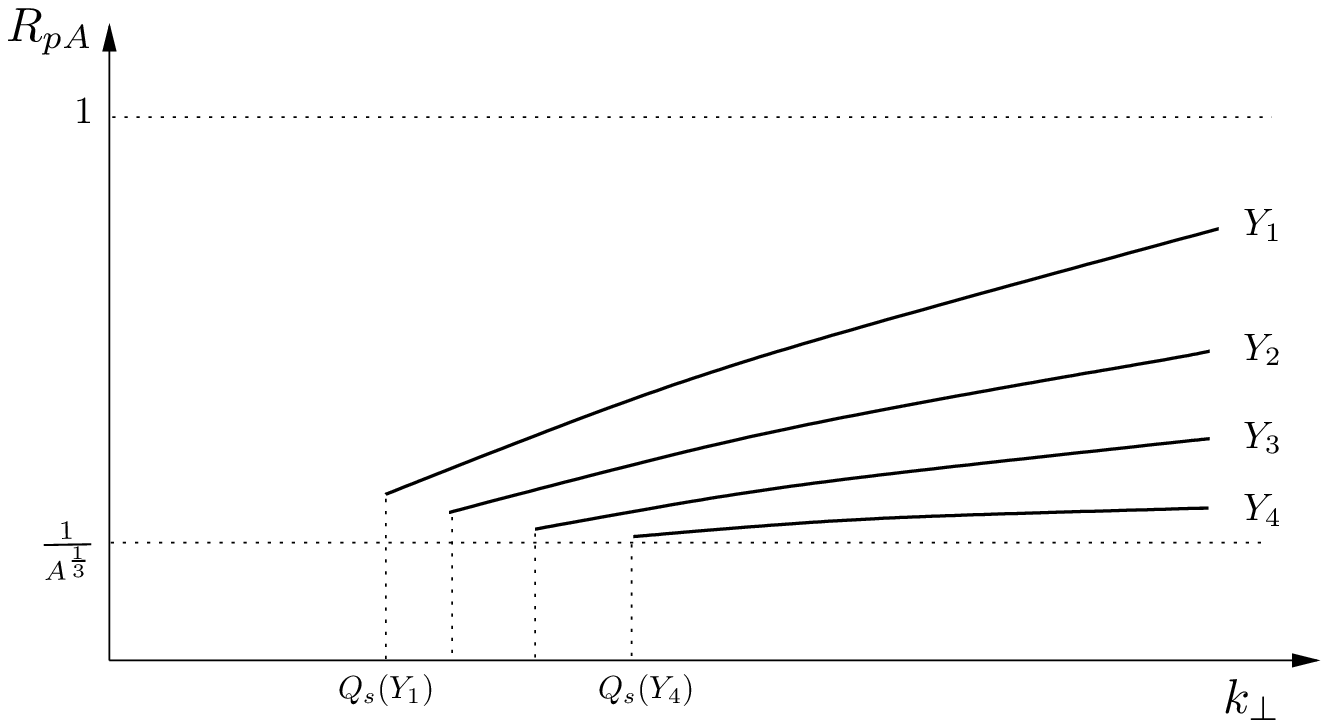, width=8cm}
\end{center}
\caption{(a) Phase diagram of a highly evolved nucleus/proton. (b) $R_{p,A}$ versus
  $k_{\perp}$ at different rapidities $Y_4 \gg Y_3\gg Y_2\gg Y_1$.} 
\label{had_wf}
\end{figure}

In the geometric scaling region shown in Fig.~\ref{had_wf}a the small-$x$
physics is reasonably described by the BK-equation which emerges in the mean
field approximation. Using the BK-equation one finds in the geometric scaling
regime in the fixed coupling case that the shape of the unintegrated gluon
distribution of the nucleus and proton as a function of $k_{\perp}$ is
preserved with increasing $Y$, because of the geometric scaling behaviour
$h_{p,A}(k_{\perp},Y)=h_{p,A}(k_{\perp}^2/Q^2_s(Y))$, and therefore the
leading contribution to the ratio $R_{pA}$ is $k_{\perp}$ and $Y$ independent,
scaling with the atomic number $A$ as $R_{pA} =1/A^{1/3(1-\gamma_{_0})}$, where
$\gamma_{_0}=0.6275$~\cite{Mueller:2003bz}. This means that gluons inside the
nucleus and proton are somewhat shadowed since $h_A/h_p = A^{\gamma_{_0}/3}$ lies
between total ($h_A/h_p=1$) and zero ($h_A/h_p=A^{1/3}$) gluon shadowing. The
{\em partial gluon shadowing} comes from the anomalous behaviour of the
unintegrated gluon distributions which stems from the BFKL evolution.

We have recently shown~\cite{Kozlov:2006qw} that the behaviour of $R_{pA}$ as
a function of $k_{\perp}$ and $Y$ in the fixed coupling case is completely
changed because of the effects of gluon number fluctuations or Pomeron loops
at high rapidity. According to~\cite{Iancu:2004es} the influence of
fluctuations on the unintegrated gluon distribution is as follows: Starting
with an intial gluon distribution of the nucleus/proton at zero rapidity, the
stochastic evolution generates an ensamble of distributions at rapidity $Y$,
where the individual distributions seen by a probe typically have different
saturation momenta and correspond to different events in an experiment. To
include gluon number fluctuations one has to average over all individual
events, $h^{fluc.}_{p,A}(k_{\perp},Y) = \langle h_{p,A}(k_{\perp},Y) \rangle$,
with $h_{p,A}(k_{\perp},Y)$ the distribution for a single event. The main
consequence of fluctuations is the replacement of the geometric scaling by a
new scaling, the diffusive scaling~\cite{Mueller:2004se,Iancu:2004es},
$\langle h_{p,A}(k_{\perp},Y) \rangle = h_{p,A}\left(\ln(k^2_{\perp}/\langle
  Q_s(Y)\rangle^2))/[D Y]\right)$. The diffusive scaling, see
Fig.~\ref{had_wf}a, sets in when the dispersion of the different events is
large, $\sigma^2 = \langle \rho_s(Y)^2 \rangle- \langle \rho_s(Y) \rangle^2 =
D Y \gg 1$, i.e., $Y \gg Y_{DS} =1/D$, where $\rho_s(Y) = \ln(Q^2_s(Y)/k_0^2)$
and $D$ is the diffusion coefficient, and is valid in the region $\sigma \ll
\ln(k^2_{\perp}/\langle Q_s(Y)\rangle^2) \ll \gamma_{_0}\, \sigma^2$. The new
scaling means that the shape of the unintegrated gluon distribution of the
nucleus/proton becomes flatter and flatter with increasing rapidity $Y$, in
contrast to the preserved shape in the geometric scaling regime. This is the
reason why the ratio in the diffusive scaling regime~\cite{Kozlov:2006qw}
\begin{equation}
R_{pA}(k_{\perp},Y) \simeq \frac{1}{A^{\frac{1}{3}\left(1-\frac{\ln A^{1/3}}{2\sigma^{2}}\right)}} \
      \left[\frac{k_{\perp }^{2}}{\langle Q_{s}(A,Y)\rangle^2}\right]^{\frac{\ln
      A^{1/3}}{\sigma^{2}}} \ 
\label{eq:A_R_DS}
\end{equation}
yields {\em total gluon shadowing}, $R_{pA} = 1/A^{1/3}$, at asymptotic
rapidity $Y$ (at fixed $A$). This result is universal since it does not depend
on the initial conditions. Moreover the slope of $R_{pA}$ as a function of
$k_{\perp}$ descreases with increasing $Y$. The qualitative behaviour of
$R_{pA}$ at fixed $\alpha_s$ due to fluctuation effects is shown in
Fig.~\ref{had_wf}b.

The above effects of fluctuations on $R_{pA}$ are valid in the fixed coupling
case and at very large energy. It isn't clear yet whether the energy at LHC is
high enough for them to become important. Moreover, in the case where
fluctuation effects are neglected but the coupling is allowed to run, a
similar behaviour for $R_{pA}$ is obtained~\cite{Iancu:2004bx}, including the
total gluon shadowing. It remains for the future to be clarified how important
fluctuation or running coupling effects are at given energy windows, e.g.,
at LHC energy.

\section*{References}
\begin{thebibliography}{10}

\bibitem{Mueller:2003bz}
  A.~H.~Mueller,
  Nucl.\ Phys.\  A {\bf 724} (2003) 223.

\bibitem{Iancu:2004bx}
  E.~Iancu, K.~Itakura and D.~N.~Triantafyllopoulos,
  Nucl.\ Phys.\  A {\bf 742} (2004) 182.

\bibitem{Kozlov:2006qw}
M.~Kozlov, A.~I.~Shoshi and B.~W.~Xiao,
arXiv:hep-ph/0612053.

\bibitem{Mueller:2004se}
  A.~H.~Mueller and A.~I.~Shoshi,
  Nucl.\ Phys.\  B {\bf 692} (2004) 175.

\bibitem{Iancu:2004es}
E.~Iancu, A.~H.~Mueller and S.~Munier,
Phys.\ Lett.\ B {\bf 606} (2005) 342.

\end{thebibliography}
\end{document}